\documentclass[aps,prl,showpacs,twocolumn,superscriptaddress,preprintnumbers,nofootinbib]{revtex4-1}

\pdfoutput=1

\usepackage{amsmath, amsfonts, amsthm, amssymb, graphicx}

\usepackage{amsmath}
\usepackage{amsfonts}
\usepackage{amssymb}
\usepackage{subeqnarray}
\usepackage{latexsym}
\usepackage[latin1]{inputenc}
\usepackage{graphicx}

\usepackage[english]{babel}

\usepackage{hyperref}
\hypersetup{ 
pdfstartview=FitV,
colorlinks=true, 
linkcolor=blue, 
citecolor=blue, 
filecolor=blue, 
urlcolor=black}

\allowdisplaybreaks[3]

\makeatletter
\@addtoreset{equation}{section}
\makeatother

\newcommand{\be}{\begin{equation}}
\newcommand{\ee}{\end{equation}}
\newcommand{\beq}{\begin{equation}}
\newcommand{\eq}{\begin{equation}}
\newcommand{\eeq}{\end{equation}}
\newcommand{\eqa}{\begin{eqnarray}}
\newcommand{\eeqa}{\end{eqnarray}}
\newcommand{\bea}{\begin{eqnarray}}
\newcommand{\eea}{\end{eqnarray}}
\newcommand{\bsea}{\begin{subeqnarray}}
\newcommand{\esea}{\end{subeqnarray}}

\newcommand{\ud}{\mathrm{d}}

\newcommand{\ka}{\kappa}
\newcommand{\la}{\lambda}

\newcommand{\hs}{{\hat{s}}}

\newcommand{\hT}{{\hat{T}}}

\newcommand{\hg}{{\hat{g}}}
\newcommand{\hG}{{\hat{G}}}

\def\hri#1#2{\href{http://arxiv.org/abs/#1}{[arXiv:#1]#2}}
\def\hre#1#2{\href{http://arxiv.org/abs/#1/#2}{[arXiv:#1/#2]}}

\begin{document}

\title{AdS/Ricci-flat correspondence  and the Gregory-Laflamme instability}

\author{Marco M. Caldarelli}
\email[]{M.M.Caldarelli@soton.ac.uk}
\affiliation{
School of Mathematics, University of Southampton,  Southampton, UK}
\affiliation{CPhT, Ecole Polytechnique, CNRS UMR 7644, Palaiseau, France}
\affiliation{LPT, Univ. Paris-Sud, CNRS UMR 8627, Orsay, France}

\author{Joan Camps}
\email[]{J.Camps@damtp.cam.ac.uk}
\affiliation{DAMTP, University of Cambridge,
Cambridge, UK}
\affiliation{Dept of Mathematical Sciences \& CPT, Durham University,
Durham, UK}
\author{Blaise Gout\'eraux}
\email[]{blaise@kth.se}
\affiliation{Nordita,
KTH Royal Institute of Technology \& Stockholm University, Stockholm, Sweden}
\affiliation{APC, Univ. Paris Diderot, CNRS/IN2P3, CEA/Irfu, Obs. de
Paris, Sorbonne Paris Cit\'e, France}
\affiliation{Crete Center for Theoretical Physics, Department of Physics,
University of Crete, Heraklion, Greece}

\author{Kostas Skenderis}
\email[]{K.Skenderis@soton.ac.uk}
\affiliation{
School of Mathematics, University of Southampton,  Southampton, UK}
\affiliation{
 Korteweg-de~Vries Institute for Mathematics \&
 Institute for Theoretical Physics, Amsterdam, Netherlands
}

\date{\today}

\begin{abstract}

We show that for every asymptotically AdS solution compactified on a torus there is a corresponding Ricci-flat solution obtained by replacing the torus by a sphere, performing a Weyl rescaling of the metric and appropriately analytically continuing the dimension of the torus/sphere (as in generalized dimensional reduction). In particular, it maps Minkowski spacetime to AdS
on a torus, the holographic stress energy tensor of AdS to the stress energy tensor due to a brane localized in the interior of spacetime and AdS black branes to (asymptotically flat) Schwarzschild black branes. 
Applying it to the known solutions describing the hydrodynamic regime in AdS/CFT, we derive the hydrodynamic stress-tensor of asymptotically flat black branes to second order, which is constrained by the parent conformal symmetry. We compute the dispersion relation of the Gregory-Laflamme unstable modes through cubic order in the wavenumber, finding remarkable agreement with numerical data. In the case of no transverse sphere, AdS black branes are mapped to Rindler spacetime and 
the second-order transport coefficients of the fluid dual to Rindler spacetime are recovered.

\end{abstract}

\pacs{04.50.Gh,11.25.Tq}
\preprint{\scriptsize CCTP-2012-25, CPHT-RR076.1112, DCPT-12/45, LPT-ORSAY 12-108, NORDITA-2012-87}

\maketitle

\paragraph{Introduction.}

Holography is very well understood when the spacetime is  asymptotically (locally) Anti-de Sitter (AdS).
However, the general arguments for holography based on black hole physics are insensitive to asymptotics, suggesting 
that one should be able to develop a holographic dictionary for spacetimes with non-AdS asymptotics.  In some cases, a string theoretic 
construction implies that a spacetime which is not asymptotically AdS should admit a holographic description.
Indeed, the original decoupling argument  \cite{Maldacena:1997re} also extends to nonconformal branes \cite{Itzhaki:1998dd} and these give rise to spacetimes that are not asymptotically AdS. 
Nevertheless, a detailed holographic dictionary can be set up for such spacetimes,
\cite{Wiseman:2008qa,Kanitscheider:2008kd}.

It turns out that these cases can be linked to the asymptotically AdS case via
 a generalized dimensional reduction \cite{Kanitscheider:2009as}:  one starts from asymptotically AdS spacetimes in higher dimensions, dimensionally reduces 
on a torus and then analytically continues the dimension of the torus (which only appears as a parameter in the reduced theory)  to a real number. This procedure commutes with all operations needed to establish a precise holographic dictionary: one can  obtain the general asymptotic solutions, the covariant boundary counterterms and the holographic 1-point functions of the reduced theory from the corresponding results of the parent AdS theory (see also  \cite{Gouteraux:2011qh} for the finite density case).

In this work we will apply generalized dimensional reduction in order to link a class of Ricci-flat solutions to a class of asymptotically AdS spacetimes. This should then provide the starting point for developing a detailed holographic dictionary for this class of Ricci-flat spacetimes, which we will
discuss in detail elsewhere \cite{upcoming}. Here we will present the AdS/Ricci-flat map and one application, namely the analytic study of the Gregory-Laflamme (GL) instability \cite{Gregory:1993vy,book}.
The GL instability is an instability of black strings  to linearized gravitational perturbations that break their translational symmetry but preserve their transverse spherical symmetry. The linearized equations were originally solved 
numerically and one of the main results is the dispersion relation of the unstable mode. Since then there has been much progress in the nonlinear evolution of the perturbation (see e.g. \cite{Lehner:2010pn, book}).
Here we will use the connection to AdS to obtain an analytic long-wavelength approximation of the dispersion relation. 
Our results reproduce the quadratic approximation of the dispersion relation found  in \cite{Camps:2010br}, extend it to 
 one order higher and lead to a  striking agreement between the analytic expression and the numerical results.

\paragraph{AdS/Ricci-flat correspondence.} 
Consider pure AdS gravity in $(d+1)$ dimensions with cosmological constant $\Lambda=-d(d-1)/2 \ell^2$
and compactify $(d-p-1)$ spatial dimensions on a diagonal torus of volume $vol(T)$. Let us  label the coordinates as $x^M = \{\rho, z^\mu\}, (\mu=0,...,d-1)$ and $z^\mu=\{x^a, \vec{y}\}, (a=0,...,p)$, with $\vec{y}$ the torus coordinates.
Then corresponding to any solution of the form,
\be
    \ud s_\Lambda^2=\ud \hs^2_{p+2}(\rho,x) + e^{2\phi(\rho,x)/(d-p-1)}\ud \vec{y}^2,
    \label{KKDiagonal}
\ee
there is a $D=n+p+3$ Ricci-flat solution given by
\be
    \ud s_0^2= e^{2\phi(\rho,x)/(n+p+1)}\left(\ud \hs^2_{p+2}(\rho,x) + \ell^2 \ud \Omega_{n+1}^2\right),
    \label{KKRicciflat}
\ee
where $\ud\Omega_{n+1}^2$ is the metric of the unit round $(n+1)$-sphere of volume $\Omega_{n+1}$ and $d$ and $n$ are related by
\be
\label{relation}
n \leftrightarrow -d\,.
\ee
In other words, if we are given a solution (\ref{KKDiagonal}) then the correspondence instructs us to 
extract from it the $(p+2)$-metric $\hat{g}$ and the scalar field $\phi$, replace in them all explicit factors of $d$ by $-n$ and then insert them in (\ref{KKRicciflat}). This metric is then Ricci-flat.\footnote{A similar, more restrictive correspondence  was noted  earlier in \cite{Charmousis:2003wm}.}
Similarly,  given (\ref{KKRicciflat}) one can follow the same steps (now with $n$ replaced with $-d$)
in order to obtain the Einstein metric (\ref{KKDiagonal}).
Note that this correspondence requires knowing the solutions for general $n$ and $d$. Furthermore, the
AdS radius $\ell$ is mapped to the radius of the sphere in (\ref{KKRicciflat}).

This correspondence is proven by substituting the Ans\"atze (\ref{KKDiagonal}) and (\ref{KKRicciflat}) in their corresponding field equations,
expressing them as equations for the $(p+2)$-dimensional metric $\hat{g}$ and the scalar field $\phi$  
and showing the equations map to each other if we use (\ref{relation}). This is another instance of generalized dimensional reduction: reducing AdS gravity on the torus to $(p+2)$ dimensions, $d$ now appears as a parameter which can be continued to negative values. This theory then matches the one obtained by starting with Einstein gravity without cosmological constant and reducing on a sphere to $(p+2)$ dimensions. 

This correspondence implies that this class of Ricci-flat solutions has an underlying  holographic structure, namely that inherited from the AdS ones. Asymptotically locally AdS spacetimes come equipped with a 
conformal structure represented by an arbitrary metric on the boundary. This metric in AdS/CFT acts as the source for the 
dual stress energy tensor.  Using the Ricci-flat/AdS correspondence, we find that this class of Ricci-flat solutions will also depend on an arbitrary metric and one may follow the usual steps to set up a holographic dictionary for Ricci-flat solutions.
This is highly nontrivial as the naive extension of the AdS holographic computations to Ricci-flat solutions reveals that the asymptotia depend on constrained data and the infinities in the on-shell action are nonlocal with respect to them \cite{deHaro:2000wj,Skenderis:2002wp}. We will restrict ourselves here to flat boundary metrics.  The derivation of the map (following closely \cite{Gouteraux:2011qh}), the generalization to curved boundaries and the full holographic analysis for this class of spacetimes will be presented in \cite{upcoming}. 

The simplest case to analyze is AdS spacetime with $(d-p-1)$ of the boundary directions compactified on a torus,
\be
\ud s_\Lambda^2 = \frac{\ell^2}{r^2} \left(\ud r^2 + \eta_{ab} \ud x^a \ud x^b + \ud \vec{y}^2\right) ,           
\ee
where $\rho=r^2$. Matching with (\ref{KKDiagonal}) we find $\phi(r,x)=(p+1-d) \ln r/\ell$ and after applying (\ref{relation}) 
and substituting in (\ref{KKRicciflat}) we obtain
\be
\ud s^2_0 = (\ud r^2 + r^2 \ud\Omega^2_{n+1}) +  \eta_{ab} \ud x^a \ud x^b,
\ee
which is simply $D$-dimensional Minkowski spacetime. On the AdS side, we fixed $\eta_{ab}$ as the boundary condition for the metric on the noncompact boundary coordinates.
This maps on the Ricci-flat side to the metric on a $p$-brane located at $r=0$,  the origin in transverse space. The radial direction in
AdS is the transverse distance from the $p$-brane. This is reminiscent of the way AdS spacetime emerges from  $D/M$-brane spacetimes in the decoupling limit \cite{Maldacena:1997re}.

 Let us now add excitations on top of AdS. The general asymptotic solution  in Fefferman-Graham (FG) gauge with
a flat boundary metric is given by \cite{de Haro:2000xn},
\be \label{FG}
\ud s_\Lambda^2 = \frac{\ud \rho^2}{4\rho^2}+\frac1\rho \left(\eta_{\mu \nu} 
+ \rho^{d/2} g_{(d)\mu\nu} + \cdots \right)
\ud z^\mu\ud z^\nu,
\ee
where here and throughout the rest of this work we set the AdS radius $\ell=1$. $g_{(d)\mu\nu}$ is  related to the expectation value of the dual stress energy tensor, \cite{de Haro:2000xn},
\be
T_{\mu \nu} = \frac{d}{ 16 \pi G_N} g_{(d)\mu \nu}\,,
\ee
which satisfies  (as a consequence of the gravitational field equations) \cite{Henningson:1998gx, de Haro:2000xn},
 \be \label{WI}
\partial^\mu T_{\mu \nu}=0\,, \qquad  T_\mu^{\phantom{1}\mu}=0\,.
\ee
These are the correct Ward identities for the CFT on a flat background.

Compactifying on a $(d-p-1)$ torus, 
the metric $\hat{g}$ and the scalar field $\phi$ can be extracted from (\ref{KKDiagonal}):
\bea
\ud \hs^2&{=}&\frac{\ud \rho^2}{4\rho^2}{+}\frac1\rho\left(\eta_{ab}{+}\rho^{d/2} (\hat{g}_{(d)ab}{+}\rho \hat{g}_{(d+2)ab}{+} \ldots) \right)\ud x^a \ud x^b, \nonumber\\
\phi &=&\frac{(p+1-d)}2 \ln \rho+\rho^{d/2} \hat{\phi}_{(d)} + \rho^{d/2+1} \hat{\phi}_{(d+2)} +  \ldots, \label{pFG}
\eea
where the displayed coefficients are related to the expectation values of the stress energy tensor 
$\hat{T}_{ab}$ and of the scalar operator $ {\cal O}_\phi$ of the $(p+2)$-dimensional theory 
\cite{Kanitscheider:2009as}:
\be
\hT_{ab}=\frac{d}{16\pi \hG_N}\hg_{(d)ab}\,, \quad \hat{\mathcal O}_\phi=-\frac{d (d-p-1)}{32 \pi \hG_N}\hat{\phi}_{(d)}\,,
\ee
while $\hat{g}_{(d+2)ab} = - \Box \hat{g}_{(d)ab}/(2 d (d+2))$ and $\hat{\phi}_{(d+2)}= - \Box \hat{\phi}_{(d)}/(2 d (d+2))$.
$\hat{G}_N=G_N/vol(T)$ is the $(p+2)$-dimensional Newton's constant. 
The stress energy tensor satisfies the expected trace and diffeomorphism Ward identities
\be 
\partial^a \hT_{ab}=0\,, \qquad \hT_a^{\phantom{1}a}=(d-p-1)\hat{\mathcal O}_{\phi}\,.
\ee

We can now apply the AdS/Ricci-flat correspondence to obtain the corresponding Ricci-flat solution:
\bea
\ud s_0^2&&=(\eta_{AB}+ h_{AB}+\dots) \ud x^A \ud x^B  = \left(1-\frac{16\pi \hG_N\, \,}{n\,r^{n}}(1+ \right.\nonumber\\
&&\left.\frac{r^2}{2 (n-2)} \Box) \hat{\mathcal O}_\phi(x) \right)
\left(\ud r^2+\eta_{ab} \ud x^a \ud x^b + r^2\ud \Omega_{n+1}^2\right) \nonumber \\&&
\  -\frac{16\pi \hG_N\, \,}{n\,r^{n}}(1+\frac{r^2}{2 (n-2)} \Box) \hT_{ab}(x) \ud x^a \ud x^b +\dots
\label{asAdSmaptoFlat}
\eea
where $\rho=r^2$ and $x^A$ are $D$-dimensional coordinates. Defining ${\bar{h}}_{AB}=h_{AB}-\frac{h^C_{\phantom{1}C}}{2}\eta_{AB}$,
\be
\Box{\bar h}_{AB}=16\pi\hG_N\Omega_{n+1}\delta_A^{\phantom{1}a}\delta_B^{\phantom{1}b} \hT_{ab}\delta^{n+2}(r)
\ee
follows through second order terms in (boundary) derivatives. 
Comparing with the linearized Einstein equations we conclude that the 
holographic stress energy tensor takes a new meaning: it is (proportional to) minus the stress energy tensor due to a $p$-brane located at $r=0$ that sources the linearized gravitational field $h_{AB}$.

 Let us now consider how black objects are mapped to each other under the 
AdS/Ricci-flat map. The planar AdS black brane reads,
\be \label{AdSBB}
\ud s_\Lambda^2 = z^2 (-f(z)d\tau^2+d\vec{x}^2+d\vec{y}^2)+\frac{\ud z^2}{z^2f(z)}\,, 
\ee
where $x^a=\{\tau, \vec{x}\}$, the $\vec{y}$ coordinates parametrize a torus, as before, and 
$ f(z)=1-1/{(b z) }^{d}$. 
 Applying the map and  changing coordinates, $z = 1/r$, we obtain
\be \label{BlackpBrane}
	\ud s_0^2=-f(r)\ud \tau^2 + \frac{\ud r^2}{f(r)}+r^2\ud \Omega_{n+1}^2+\ud \vec{x}^2,
\ee
where $ f( r)=1-(b/{r})^{n}$, i.e. the Schwarzschild black $p$-brane.  

\paragraph{Gregory-Laflamme instability.} Black strings and branes suffer from a linearized s-wave instability 
\cite{Gregory:1993vy}. Most of our results are applicable for $p$-branes but 
 for concreteness we will focus on the black string case, $p=1$. Since the perturbations respect the transverse spherical symmetry, the perturbed metrics fall within 
the class of metrics (\ref{KKRicciflat}). This means that we can use the AdS/Ricci-flat correspondence to map them
to linear perturbations of the AdS black brane (\ref{AdSBB}). 
These perturbations are known analytically for small frequencies and we will use this result to obtain an approximate analytic formula for the dispersion relation of the GL mode.

The GL instability is a long wavelength instability: only modes with wavelength greater than a critical value are unstable
and the critical value is related to the thickness of the brane $b$. Thus, compactifying the $x$ coordinate on a small enough 
circle may stabilize the black string. Here we consider the case $x$  is noncompact. In this  case,  unstable modes appear below a critical wavenumber $k_c$. Their   (imaginary) frequency $\Omega$  and momentum $k$ satisfy a dispersion relation $\Omega(k)$, which we would like to compute.

We will analyze the problem in the small $k$ limit.
This regime is mapped under the AdS/Ricci-flat correspondence to the gravitational solutions describing the hydrodynamic regime of the dual CFT. These solutions were obtained in  \cite{Bhattacharyya:2008jc,Bhattacharyya:2008mz,book} by solving the full (nonlinear) Einstein equations in a (boundary) derivative expansion. The AdS/Ricci-flat correspondence then leads to  
the construction of similar nonlinear perturbations of the black string, which 
may shed light on
the nonlinear GL instability. These (nonlinear) solutions to second order in gradients will be presented in \cite{upcoming}. Here we will linearize in the amplitude of perturbations and discuss their dispersion relation.

The gravitational solution is essentially encoded on the dual stress energy tensor $T_{\mu \nu}$. On the AdS side, 
this is the statement of  holography.  On the Ricci-flat side, we saw that $\hat{T}_{ab}$ sources the linearized gravitational 
field.  Thus, to determine the dispersion relation, it suffices to solve the Ward identities (\ref{WI}). This provides also a 
direct link with the blackfold formalism \cite{Emparan:2009cs,Emparan:2009at,Camps:2012hw,book}, which was already used in order to study the 
GL instability \cite{Camps:2010br}.

Consider first the equilibrium solutions. On the AdS side, the
solution is dual to an ideal conformal fluid. The equation of state follows from \eqref{WI}
\be
 \varepsilon=(d-1)P
\ee
where $\varepsilon$ and $P$ are the energy and pressure densities. Using the AdS/Ricci-flat map one finds,
 \be \label{eqS}
 \tilde{\varepsilon}= -(n+1)\tilde{P}\,,\qquad c_s^2=\frac{\partial \tilde{P}}{\partial \tilde{\varepsilon}}=-\frac1{n+1}\,,
\ee
where $\tilde{\varepsilon}$ and $\tilde{P}$ are the ADM energy and pressure densities. 
Since the speed of sound $c_s$ is imaginary, there is an instability for the sound modes which is linked to the GL instability \cite{Emparan:2009at}.

The holographic stress energy tensor for gravitational solutions describing the hydrodynamic regime has been obtained
for general $d$ and to second order gradients in \cite{Bhattacharyya:2008mz}. 
Defining the $d$-velocity $u^\mu=\{u^a,\vec{0}\}$ with components only in the noncompact directions ($u^\mu u_\mu=-1$), the projection operator $P_{\mu\nu}=\eta_{\mu\nu}+u_\mu u_\nu$ orthogonal to $u_\mu$ and the shear tensor $\sigma_{\mu\nu}=P_\mu^\ka P_\nu^\la \partial_{(\ka}u_{\la)}-P_{\mu\nu}\partial\cdot u/(d-1)$,
the terms contributing to the linearized analysis are 
\be \label{tmunu}
\begin{split}
{T}_{\mu\nu} =& {P}(\eta_{\mu\nu}+ d{u}_\mu {u}_\nu)-2{\eta}{\sigma}_{\mu\nu} \\
	&\qquad-2{\eta}({\tau}_\omega-b) {P}_\mu{}^\ka{P}_\nu{}^\la{u}\cdot\partial {\sigma}_{\ka\la}
\end{split}
\ee
with
\be
P=\frac{b^{-d}}{16 \pi G_N}\,,\quad\eta =\frac{b^{1-d}}{16 \pi G_N}\,, \quad \tau_{\omega} 
= -\frac{b}{d} H_{2/d-1}\,,
\ee
giving the pressure, shear viscosity $\eta$ and 
second order coefficient  $\tau_\omega$. $H_z$ is the Harmonic number function. Note that $H_z$ is an analytic function in $z$, defined over the whole complex $z$-plane with  singular points at all negative integers.

For any solution of the relativistic Navier-Stokes equation, $\partial_\mu T^{\mu \nu}=0$, with $T_{\mu \nu}$ as above, 
there is a corresponding gravitational solution. To obtain the dispersion relation for the sound mode we consider an infinitesimal variation of the velocity in the $x$ direction, $u^x \sim \delta u^x \exp (-i \omega t + i k x)$ and of the
energy density $\varepsilon \sim \varepsilon_0+ \delta \varepsilon \exp (-i \omega t + i k x)$, where $\varepsilon_0$ is the equilibrium energy density.  Applying the AdS/Ricci-flat correspondence at the level of solutions and following \cite{Baier:2007ix} leads to the 
dispersion relation for the sound mode of the black string to order $k^3$. Defining $\Omega=-i \omega$ we find\footnote{Note that the naive application of the map directly in the dispersion  relation of \cite{Baier:2007ix}
would lead to incorrect sign of the $k^2$ term. This and other such subtleties will be explained in \cite{upcoming}.}
\beq
\Omega=\frac{1}{\sqrt{n{+}1}}k-\frac{(2{+}n)}{n(1{+}n)}k^2+\frac{(2{+}n)[2{+}n(2\tilde{\tau}_\omega{-}1)]}{2n^2(1{+}n)^{3/2}}k^3, 
\label{disprel}
\eeq
where 
$
\tilde{\tau}_{\omega} = (1/n) H_{-2/n-1}
$
and we have set $b=1$. 
The first two terms reproduce the quadratic approximation in \cite{Camps:2010br}.

As mentioned above, $H_z$  is well-defined when its index is a nonnegative integer, so 
the dispersion relation (\ref{disprel}) is valid for all cases, except when $n=1,2$,\footnote{For $n \to \infty$, $H_{-2/n-1}$ is singular but $\tilde{\tau}_\omega \to 1/2$.} corresponding to 
five and six-dimensional black strings.  These cases are subtle because under (\ref{relation}) they map to $d=-1,-2$, and terms in the FG expansion (\ref{FG}) that are normally distinct now become degenerate: for 
example, 
when $d=-2$, the pairs $\{g_{(0)}, g_{(d+2)}\}$  and $\{g_{(d)}, g_{(2 d+2)}\}$. 
There are similar degeneracies for any integer $n$,  but this subtlety becomes 
relevant only  at specific order in the derivative expansion, which is $n$ ($n+1$), if $n$ is even (odd). Thus, if one 
works to second order in derivatives only the $n=1,2$ cases have to be analyzed separately. 
The complete analysis (see \cite{upcoming}) leads to
\bea
n=1 &:& \quad \Omega= \frac{1}{\sqrt{2}}k-\frac{3}{2}k^2+\frac{75}{16\sqrt{2}}k^3+O(k^4)\,, \\
n=2 &:& \quad
\Omega=\frac{1}{\sqrt{3}}k-\frac{2}{3}k^2+\frac{5}{6\sqrt{3}}k^3+O(k^4)\,.
\eea

\begin{figure}
	\centering
	\begin{tabular}{c}
	\includegraphics[width=.47\textwidth]{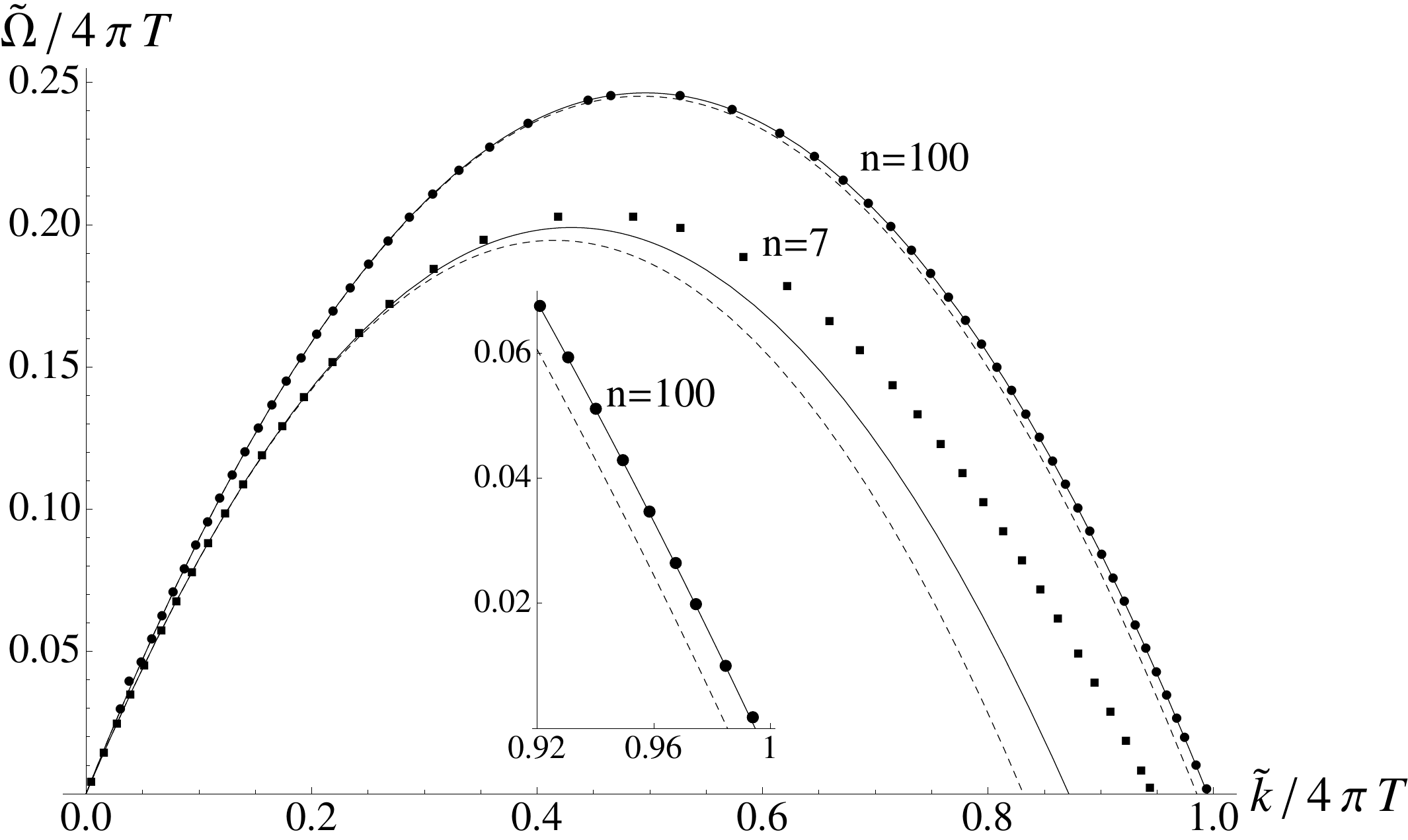}
	\end{tabular}
	\caption{\label{Fig1}Gregory-Laflamme dispersion relation for $n=7$ and $n=100$. The dots/squares represent numerical data
for $n=100/n=7$, 
the solid line 
the cubic approximation and the dashed line the quadratic approximation. In the insert we zoom into the region close to the threshold mode ($n=100$).}
\end{figure}

To obtain (\ref{disprel}) we assumed
that $k$ is small and in general one should not expect that 
it accurately captures the entire range of unstable modes, $k \leq k_c$. Nevertheless, the dispersion relation
(\ref{disprel}) agrees remarkably well with the numerical data. In Figure \ref{Fig1} we plot \eqref{disprel}  against the numerical data \cite{Figueras}  for two cases $n=7$ and $n=100$ ($\tilde{k}$ and $\tilde{\Omega}$ are defined below in (\ref{tilde})).
When $n=100$, the agreement all the way up to the threshold mode is striking.  In the same graph we also include the quadratic approximation to illustrate the improvement due to the $k^3$ term.

As it is clear from the $n=7$ case, for low $n$ the agreement becomes less and less good as $k$ becomes larger and larger. For $n=7$, (\ref{disprel})  estimates the largest $\Omega$ with an error of $2\%$, and the intercept $k_c$ is off by $9\%$. For $n=100$, the discrepancy is $0.01\%$ at the maximum and  $0.02\%$ at the intercept. For very low $n$ ($n <5$) the quadratic approximation is actually better than the cubic one when $k$ is large enough, indicating that (\ref{disprel}) is actually an asymptotic series rather than a convergent one. When $n=1, 2$ the cubic 
approximation no longer captures the correct finite $k$ behavior -- there is no threshold mode. 

To understand better the large $n$ behavior,\footnote{The GL instability in a large number of dimensions was first studied in \cite{Kol:2004pn}.} let us zoom in this region by
scaling the frequency and wavenumber as \cite{Camps:2010br},
\be \label{tilde}
	\tilde\Omega=n\, \Omega\,,\qquad \tilde k=\sqrt{n}\,k\,,
\ee
while keeping the temperature, $T=n/(4 \pi b)$, fixed.
The relative scaling of $\Omega$ to $k$ follows from the fact that the speed of sound goes to zero as $1/\sqrt{n}$  in this limit,
and the overall scaling is obtained by requiring that the dispersion relation has a nontrivial limit.
In terms of these variables the dispersion relation, truncated at $\tilde{k}^3$, takes the following suggestive form
\beq\label{DispRelLargeN}
\tilde{\Omega}=\tilde{k}\left(1-\frac{\tilde{k}}{4\pi T}\right)-\frac{\tilde{k}}{n}\left(\frac{1}{2}+\frac{\tilde{k}}{4\pi T}-\frac{\tilde{k}^2}{(4\pi T)^2}
\right) +O\left(\frac{1}{n^2}\right).  
\eeq
Solving for the threshold mode from \eqref{DispRelLargeN}, we find that $\tilde{k}_c =4\pi T(1 -1/(2n)+ O(1/n^2))$, in precise agreement with \cite{Asnin:2007rw}.
This means that $k_c/T \sim 4 \pi/\sqrt{n} \to 0$, as $n \to \infty$, and therefore the threshold mode falls within the long-wavelength approximation.

\paragraph{The fluid dual to vacuum Einstein gravity.}

In the special case $n=-1$,\footnote{When $n\to-1$, the sphere collapses in (\ref{BlackpBrane}): changing coordinates to $\left(t_R= t/b + \log f(r),\, r_R=b^2 f(r)\right)$, the metric becomes Minkowski spacetime in ingoing Rindler coordinates. } we are describing hydrodynamic perturbations of Rindler spacetime, whose fluid dual was studied in \cite{Bredberg:2011jq, Compere:2011dx, Compere:2012mt, Eling:2012ni}. In particular, this fluid has zero equilibrium energy density $\tilde{\varepsilon}$ and it is characterized by its pressure density $\tilde{P}$. Indeed, as $n \to -1$ with $\tilde{P}$ fixed we find that (\ref{eqS}) yields $\tilde{\varepsilon}=0$. All transport coefficients through second order may be obtained by reducing the results of  \cite{Bhattacharyya:2008mz} to $p+2$ dimensions (as described earlier around (\ref{tmunu}) but without linearizing) and converting to the isotropic gauge used in \cite{Compere:2012mt}. Taking the limit $n \to -1$ leads to exact agreement with \cite{Compere:2011dx,Compere:2012mt, Eling:2012ni}.

\paragraph{Conclusions.} We presented a map between a class of asymptotically AdS spacetimes and Ricci-flat solutions.
This map allowed us to construct and infer properties of Ricci-flat solutions from corresponding AdS ones. In particular,  we 
studied the GL instability and  obtained more insight about the Rindler fluid. The essential features of both the GL instability (i.e. that there is an instability) and the Rindler fluid (i.e. that the energy density is zero) follow from the conformal equation of state  of the parent theory, $ \varepsilon=(d-1)P$,  upon use of the map. In the first case, the map leads to an imaginary speed of sound signaling an instability  and in the second case to zero energy density. The detailed form of the transport properties is also dictated by the parent theory and as such it is constrained by conformal symmetry.

A major motivation for this work and one of the future research directions is setting up holography for Ricci-flat spacetimes. 
The AdS/Ricci-flat map should allow us to develop a detailed holographic dictionary for the class of Ricci-flat spacetimes discussed here, including the case of AdS curved boundaries respecting the ansatz \eqref{KKDiagonal}. This ansatz can capture deformations of the transverse sphere on the Ricci-flat side which preserve a round sub-sphere. To go beyond this and capture more general asymptotically flat spacetimes, we would need to turn on arbitrary source terms on the AdS side (such terms were set to zero here)
and consider deformations of the internal manifolds on either side of the map. It would also be interesting to include matter fields on either side.

 The map also implies that there is an inherited (hidden) conformal invariance on the 
Ricci-flat side. For example, this generalized conformal structure \cite{Kanitscheider:2008kd} dictates the form of the transport coefficients for nonconformal branes \cite{Kanitscheider:2009as} and related systems
\cite{Gouteraux:2011qh} and (using the results in this work) also explains the form of the first order 
transport coefficient in \cite{Camps:2010br}. In recent black hole literature a hidden conformal invariance was found in a number of
instances, see, for example, \cite{Castro:2010fd,Bertini:2011ga,Cvetic:2011hp,Carlip:2011vr}.  It would be interesting to understand if there is a
link between these works and the conformal symmetry inherited from AdS via the AdS/Ricci-flat map.
We intend to pursue these and related questions in the near future.

\paragraph{Acknowledgements.} 
We are indebted to P. Figueras for kindly providing the numerical data for the Gregory-Laflamme instability used in Figure \ref{Fig1}. We would like to thank R. Emparan for many discussions and valuable comments on the draft, as well as bringing to our attention \cite{Asnin:2007rw} and pointing out the agreement of our result for the threshold mode with the order $O(1/n)$ result there.
We would also like to thank C. Charmousis, V. Hubeny, D. Klemm, M. Rangamani, 
S. Ross and B. Withers for discussions. 
JC acknowledges support from  the STFC Consolidated Grant ST/J000426/1 and the European Research Council grant no. ERC-2011-StG 279363-HiDGR and would like to 
thank the institutes in Orsay, Polytechnique, Carg\`{e}se, APC, for hospitality during the course of this work.
JC and BG would like to thank the program ``The holographic way: string theory, gauge theory and black holes" held at NORDITA for a stimulating environment during the final stages of this work. 
MC, JC,  BG would like to thank the Institute of Physics in Amsterdam for hospitality. 
MC acknowledges support from  ANR, ERC Advanced Grant, LABEX
P2IO and ITN grants.
KS acknowledges support from NWO via a VICI grant.
MC and KS acknowledge support from  the John Templeton Foundation.
This publication was made possible through the support of a grant from the 
John Templeton Foundation. The opinions expressed in this publication are those of the authors and do not necessarily reflect the views of the John Templeton Foundation.

\end{document}